\newcommand{\lsim}{\mathrel{\rlap{\lower4pt\hbox{\hskip1pt$\sim$}}
                   \raise1pt\hbox{$<$}}}
\newcommand{\gsim}{\mathrel{\rlap{\lower4pt\hbox{\hskip1pt$\sim$}}
                   \raise1pt\hbox{$>$}}}

\documentstyle[12pt]{article}
\begin{document}

\title{Mutated Hybrid Inflation}
\author{Ewan D. Stewart\thanks{e-mail address:
eds@murasaki.scphys.kyoto-u.ac.jp} \\ Department of Physics \\
Kyoto University \\ Kyoto 606, Japan}
\maketitle
\begin{abstract}
A new model of inflation is described. An unusual form for the
inflationary potential is obtained because the inflaton corresponds
to a non-trivial path in the configuration space of the two real
scalar fields of the model. The model predicts a spectral index
$n = 1 - 3/2N \simeq 0.97$ for the density perturbations and
negligible gravitational waves.
\end{abstract}
\vspace*{-70ex}
\hspace*{\fill}{\bf KUNS 1275}\hspace*{2.3em}\\
\hspace*{\fill}{astro-ph/9407040}
\thispagestyle{empty}
\setcounter{page}{0}
\newpage
\setcounter{page}{1}

\section{Introduction}
The model of inflation \cite{i} discussed in this paper is closely
related to Linde's False Vacuum (or `Hybrid') Inflation idea
\cite{fvi}, and shares the important property that chaotic inflation
occurs for values of the inflaton field well below the Planck scale.
However, unlike the usual version of False Vacuum Inflation, this
model has a non-trivial inflationary trajectory [Eq.~(\ref{traj})]
which leads to an unusual form for the inflationary potential
[Eq.~(\ref{pot})].

I set $ m_{\rm Pl}^{2}/8\pi = 1 $ throughout the paper.

\section{The Model}
The effective potential of the model is
\begin{equation}
\label{V}
V = \frac{1}{2} m^2 \left( \psi - \sqrt{2}\, M \right)^2
	+ \frac{1}{4} \lambda^2 \phi^2 \psi^2
\end{equation}
where $\psi$ and $\phi$ are real scalar fields with canonical kinetic
terms. I will assume $m \ll \lambda \lsim 1$ and $M \lsim 1$.

For chaotic initial conditions $ \lambda^2 \phi^2 \psi^2 / 4 $
will initially be the dominant term giving effective masses
$ m_\phi = \lambda \psi / \sqrt{2} $ and
$ m_\psi = \lambda \phi / \sqrt{2} $.
Thus if initially $ \phi > \psi $, {\mbox i.e.} half the initial
condition space, $\psi$ will decrease much more rapidly than $\phi$
and so the fields will rapidly approach the inflationary
trajectory $ \psi \phi^2 = 2 \sqrt{2}\, m^2 M / \lambda^2 $ with
$ \phi^2 \gg m/\lambda \gg \psi $ described below.
Henceforth I will assume that $\phi \ll 1$ because for
$\phi \gsim 1$ higher order terms in $\phi$ will become important.

It will be convenient to define the quantity
\begin{equation}
\alpha(\phi) \equiv \frac{ 2 m^2 }{ \lambda^2 \phi^2 }
\end{equation}
We will see [Eq.~(\ref{i})] that inflation occurs for
$\phi^2 \gg \alpha$ and so, as we are assuming $\phi \ll 1$, we see
that $\alpha \ll 1$ during inflation.

The potential Eq.~(\ref{V}) can be rewritten as
\begin{equation}
V = \frac{ m^2 M^2 }{ 1 + \alpha }
+ \frac{1}{4} \left( 1 + \alpha \right) \lambda^2 \phi^2
\left( \psi - \frac{ \sqrt{2}\, \alpha M }{ 1 + \alpha } \right)^2
\end{equation}
Now provided $ \alpha \ll 1 $ and $ M \lsim 1 $, then $\psi$ will be
constrained to its minimum for fixed $\phi$,
\begin{equation}
\label{traj}
\psi = \frac{ \sqrt{2}\, \alpha M }{ 1 + \alpha }
	\simeq \frac{ 2 \sqrt{2}\, m^2 M }{ \lambda^2 \phi^2 }
\end{equation}
We then get the inflationary potential
\begin{equation}
\label{pot}
V  = \frac{ m^2 M^2 }{ 1 + \alpha }
\simeq m^2 M^2 \left( 1 - \frac{ 2 m^2 }{ \lambda^2 \phi^2 } \right)
\end{equation}
and the non-minimal kinetic terms
\begin{equation}
\frac{1}{2} \left( 1 + \frac{ 8 \alpha^2 M^2 }{ \phi^2 } \right)
	\left( \partial \phi \right)^2
\end{equation}
The non-minimal kinetic terms arise because the inflaton is a
combination of $\phi$ and $\psi$. However, we will see [Eq.~(\ref{i})]
that the second term in the brackets is small during inflation and so
I will neglect it giving canonical kinetic terms. Now
\begin{equation}
\label{Vprime}
\frac{V'}{V} \simeq \frac{2\alpha}{\phi} \;\;\;\;{\rm and}\;\;\;\;
\frac{V''}{V} \simeq - \frac{6\alpha}{\phi^2}
\end{equation}
and we see that inflation occurs for
\begin{equation}
\label{i}
\phi^2 \gg \alpha
\end{equation}
The number of $e$-folds until the end of inflation is given by
\begin{equation}
\label{N}
N \simeq \int \frac{V}{V'} d\phi \simeq \frac{\phi^2}{8\alpha}
\end{equation}
Note that our assumption $ \phi \ll 1 $ now implies
\begin{equation}
\label{m}
m \ll \frac{ \lambda }{ 4 \sqrt{N} }
\end{equation}
The spectral index of the density perturbations is given by
\begin{equation}
n \simeq 1 - 3 \left( \frac{V'}{V} \right)^2 + 2 \frac{V''}{V}
	\simeq 1 - \frac{3}{2N} \simeq 0.97
\end{equation}
The ratio of gravitational waves to density perturbations is
given by
\begin{equation}
R \simeq 6 \left( \frac{V'}{V} \right)^2
	= \frac{ 3 m }{ 2 \lambda \left( N \right)^{3/2} }
	\ll \frac{3}{ 8 N^2 } \sim 10^{-4}
\end{equation}
where I have used Eq.~(\ref{m}). The COBE normalisation
\cite{cobe} gives
\begin{equation}
\frac{ V^{3/2} }{V'} = 2 N^{3/4} \sqrt{ \lambda m }\, M
	= 6 \times 10^{-4}
\end{equation}
For example, setting $M=1$ and $ m = \Lambda^2 $ gives
$ \Lambda = 4 \lambda^{-1/2} \times 10^{13}\,$GeV.
Also Eq.~(\ref{m}) now gives $V^{1/4} \ll 4 \times 10^{15}\,$GeV.

\section{Particle Physics Motivation}
This model of inflation might arise from a superpotential of the form
\begin{equation}
W = \Lambda^2 f(\Psi) \Xi_1 + \lambda \Phi \Psi \Xi_2
\end{equation}
where $\Phi$, $\Psi$, $\Xi_1$ and $\Xi_2$ are complex scalar fields.
The first term might be derived from some non-perturbative mechanism
such as gaugino condensation which dynamically generates the scale
$\Lambda \ll 1$, while the second term would already be present at
tree-level. This superpotential would then lead to the potential
\begin{equation}
V = \Lambda^4 \left| f(\Psi) \right|^2 + \lambda^2 |\Phi|^2 |\Psi|^2
\end{equation}
with the $\Xi$ fields constrained to zero. Without loss of generality,
taking $ f(0) = 1 $ and $ f'(0) = -a < 0 $, and expanding about
$\Psi=0$, gives
\begin{equation}
V = \Lambda^4 \left( 1 - 2a\, {\rm Re}\, \Psi + \ldots \right)
	+ \lambda^2 |\Phi|^2 |\Psi|^2
\end{equation}
During inflation, the $ \lambda^2 |\Phi|^2 |\Psi|^2 $ term will
dominate the terms quadratic in $\Psi$ derived from $f$. Then setting
$ \phi = \sqrt{2}\, |\Phi| $ and $ \psi = \sqrt{2}\, {\rm Re}\, \Psi $
we get Eq.~(\ref{V}) \footnote{Except for the $ m^2 \psi^2 / 2 $
term which is negligible during inflation and is just added to give
the model a minimum with zero potential energy in as simple a way as
possible.} with $ M = 1/a $ and $ m = a \Lambda^2 $.
If $\Lambda$ corresponds to the scale at which the hidden sector gauge
group that leads to gaugino condensation \cite{g} becomes strong, then
one would expect $\Lambda \sim 10^{13} - 10^{14}\,$GeV.

\subsection*{Acknowledgements}
I thank D. H. Lyth, M. Sasaki and the referee for helpful comments on
the draft of this paper.
I am supported by a JSPS Postdoctoral Fellowship and this work was
supported by Monbusho Grant-in-Aid for Encouragement of Young
Scientists No.\ 92062.

\end{document}